\documentclass[prb,aps,twocolumn,superscriptaddress,longbibliography]{revtex4-1}
\usepackage{graphicx,color}
\usepackage{amsthm}
\usepackage{amsfonts}
\usepackage{algorithmic}
\usepackage{enumerate}
\usepackage{latexsym}
\usepackage{amsmath}
\usepackage{amssymb}
\usepackage{graphicx}
\usepackage[colorlinks=true,citecolor=blue,linkcolor=blue]{hyperref}

\emergencystretch\maxdimen
\begin{document}

\title{Zigzag edge ferromagnetism of triangular-graphene-quantum-dot-like system}
\author{Runze Han}
\affiliation{Department of Physics, Beijing Normal University, Beijing 100875, China\\}
\author{Jiazhou Chen}
\affiliation{Department of Physics, Beijing Normal University, Beijing 100875, China\\}

\author{Mengyue Zhang}
\affiliation{Department of Physics, Beijing Normal University, Beijing 100875, China\\}

\author{Jinze Gao}
\affiliation{Department of Physics, Beijing Normal University, Beijing 100875, China\\}

\author{Yicheng Xiong}
\affiliation{Department of Physics, Beijing Normal University, Beijing 100875, China\\}

\author{Yue Pan}
\email{pany@mail.bnu.edu.cn}
\affiliation{Department of Physics, Beijing Normal University, Beijing 100875, China\\}

\author{Tianxing Ma}
\email{txma@bnu.edu.cn}
\affiliation{Department of Physics, Beijing Normal University, Beijing 100875, China\\}
\affiliation{Key Laboratory of Multiscale Spin Physics (Ministry of Education), Beijing Normal University, Beijing 100875, China\\}

\begin{abstract}
Here, the magnetic susceptibility of a triangular-graphene-quantum-dot-like system was examined by using the determinant quantum Monte Carlo method.
We focused on three zigzag edge quantum dots or rings, namely, the triangular graphene quantum ring, bilayer triangular graphene quantum dot, and bilayer triangular graphene quantum ring. The triangular-graphene-quantum-dot-like system exhibited robust edge ferromagnetic behavior, which was independent of size, monolayer or bilayer, or dot or ring shape, according to the numerical results.
At half filling, the edge magnetic susceptibility is increased by on-site interactions, especially in the low-temperature region.
Spintronics systems may benefit from use of this system due to its robust edge ferromagnetic behavior.
\end{abstract}
\maketitle

\section{Introduction}
The rich physical features of graphene have led to its extensive research and application development in the domains of electrical \cite{doi:10.1126/science.1102896,RevModPhys.81.109,PhysRevB.107.115140,Fan_2015}, optics \cite{PhysRevB.106.035401,doi:10.1126/science.1231119}, and other fields over the past few decades.
Numerous research groups have explored the properties which were effected by the stacking mode in multilayer graphene \cite{wang2021single,son2016hydrogenated,shen2020correlated,
pantaleon2023superconductivity,Nimbalkar2020,yi2021magic, PhysRevLett.97.036803,murata2019high,PhysRevB.107.L121405,
PhysRevLett.99.256802,cao2018unconventional,lu2019superconductors}. Additionally, with the increasing demand for low-power devices, the field of spintronics is rapidly developing, and novel magnetic properties have also been found in both monolayer and bilayer graphene. Therefore, the investigation into the possible ferromagnetism of graphene is significant for expanding its use in spintronic applications.
	
The electronic and magnetic properties of various graphene systems are significantly influenced by their edge atomic configurations \cite{HUANG2019310,alimohammadian2020observation,pixley2019ferromagnetism,PhysRevLett.100.047209,PhysRevB.54.17954}, such as armchair or zigzag types. Zigzag-edge graphene exhibits magnetism due to ferromagnetic coupling along each zigzag edge and antiferromagnetic coupling between two parallel zigzag edges. The presence of strong ferromagnetic coupling along zigzag edges has been theoretically predicted \cite{zhang2022ferromagnetism,doi:10.1126/science.aaw3780,PhysRevB.94.075106} and verified by experiments \cite{ruffieux2016surface,blackwell2021spin,sun2017magnetism}. The disruption of sublattice symmetry by zigzag edges is a primary factor contributing to graphene's magnetism \cite{PhysRevB.98.115428}. Furthermore, experimental studies have observed spin-related phenomena to arise from zigzag edges in graphene \cite{tada2012spontaneous,tada2013electron}. Since zigzag edges as effective strategies have been attempted to realize ferromagnetic ordering, graphene nanoribbons (GNRs) \cite{Li_2016,PhysRevB.107.125411,10.1088/978-0-7503-1701-6} and graphene quantum dots (GQDs) \cite{PhysRevB.105.235415,hu2019room,das2016size} have received increasing attention. Both of these can be thought of as putting constraints on an endlessly long two-dimensional lattice, and the edge region itself shows a variety of magnetic phenomena. GNRs allow for infinite extension in one direction while being finite in another. However, GQDs are nanometric in all dimensions and display remarkable optoelectronic properties due to quantum confinement and edge effects, as compared to other quantum dots \cite{das2016size}. Theoretical approaches, like quantum Monte Carlo (QMC) method and density functional theory (DFT) simulations \cite{das2016size,hu2019room}, have predicted strong edge magnetism even in GQDs. Recent advancements in fabrication techniques have promoted the precise creation of GQDs with varied shapes and sizes, offering a unique opportunity to investigate the impact of zigzag edges on GQDs' magnetic properties.
	
Among various graphene quantum dot structures, triangular-graphene-quantum-dot-like (TGQD-like) systems stand out, including triangular graphene quantum dots (TGQDs) \cite{wang2008graphene,hrivnak2022electron}, bilayer TGQDs \cite{mirzakhani2023magnetism}, triangular graphene quantum rings (TGQRs) \cite{PhysRevB.83.174441,PhysRevB.79.125414}, and bilayer TGQRs \cite{PhysRevB.105.115430}. Although the magnetic properties of TGQDs consistent with Lieb's theorem where boundary conditions influence energy spectra in finite-sized systems, magnetic fields do not impact edge states \cite{PhysRevB.84.245403,PhysRevLett.99.177204,tiutiunnyk2023electronic,PhysRevLett.62.1201}. Moreover, experimental efforts have significantly advanced in probing the frontier molecular orbitals of TGQDs \cite{Pavlicek2017-ji,doi:10.1021/jacs.9b05319,doi:10.1126/sciadv.aav7717,Li2021-rs}. As we all know, stacking layers of graphene can have a significant impact on its magnetic properties. The electronic and transport properties of bilayer graphene quantum dots has also been recently reported \cite{PhysRevB.101.075413,PhysRevX.8.031023,PhysRevB.94.035415,ge2020visualization}. However, the geometry of TGQD-like systems make such systems harder to be studied by analytical methods, necessitating the use of numerically exact methods for investigating TGQDs and TGQRs with zigzag edges.
	
In this work, we will further provide numerical simulations on the magnetism of the Hubbard model in TGQD-like systems using the determinant quantum Monte Carlo (DQMC) method. We observe that the edge magnetic susceptibility of three types of quantum dots (rings) at finite temperatures exhibits Curie-Weiss behavior, indicating the edge ferromagnetism's robustness in TGQD-like systems, regardless of size, layering, and shape. Notably, at low temperatures, the edge magnetic susceptibility increases with the on-site Hubbard interaction near the half-filling state. This robust edge ferromagnetic behavior holds potential implications for spintronic applications.

\section{Model and methods}

\begin{figure}[tbp]
\includegraphics[scale=0.4]{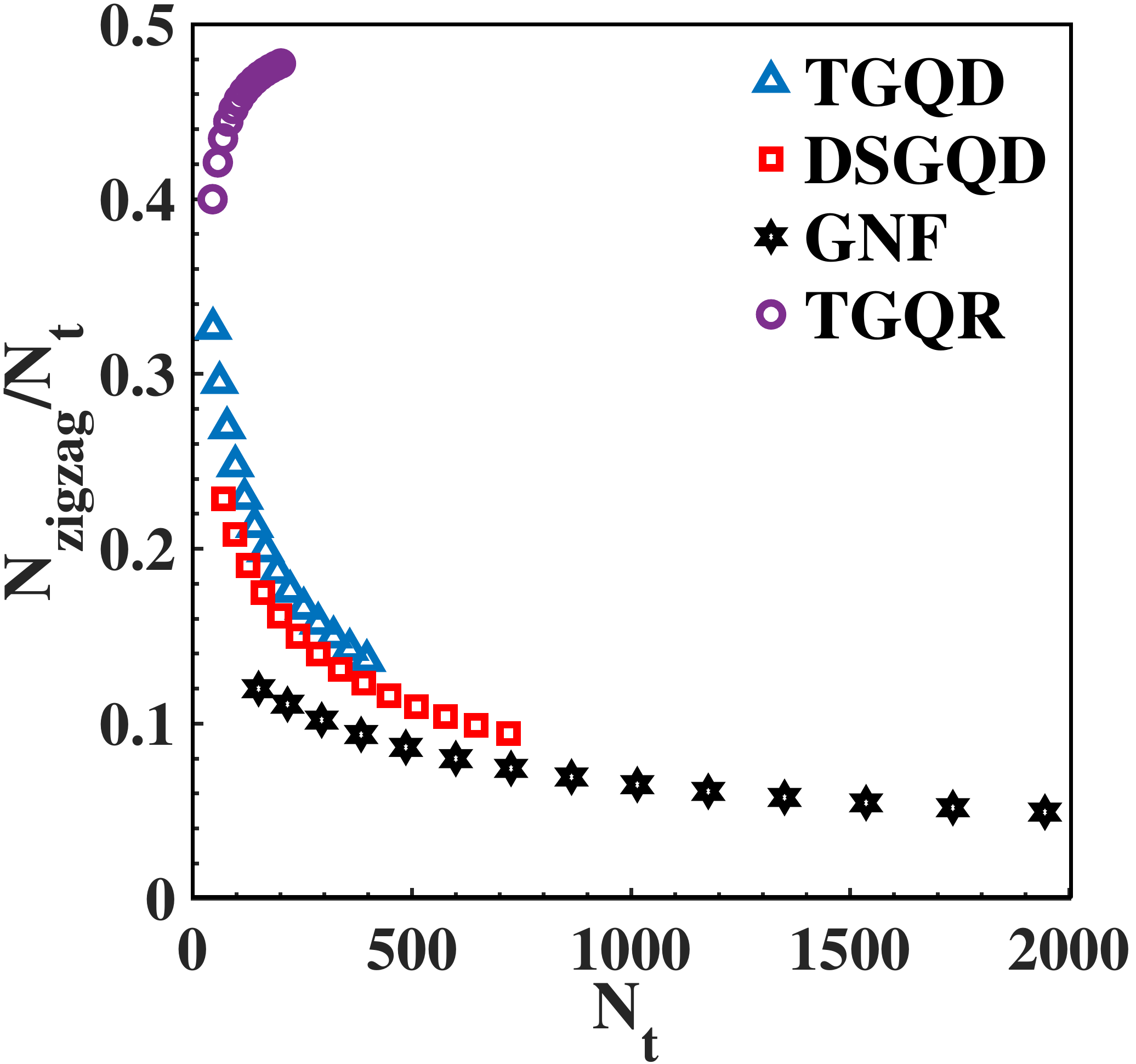}
\caption{The relationship between the ratio of the number of atoms at zigzag edge to total number of atoms $N_{zigzag}/N_{t}$ and $N_{t}$ of TGQD, diamond shaped GQD (DSGQD), graphene nanoflake (GNF), TGQR, respectively.
}
\label{relation_ratio}
\end{figure}

We select the quantum dot model that most accurately captures the zigzag edge circumstances because the zigzag edge of graphene
is more prone to ferromagnetism.
We determined that the total number of atoms $N_t$ varies with the percentage $N_{zigzag}/N_t$ of graphene nanoflakes,
diamond-shaped GQDs, TGQDs, and TGQRs with zigzag edges at different lattice sites, as shown in Fig. \ref{relation_ratio}.

It is apparent that the ratio of edge atoms in quantum rings is greater than that in bulk quantum dots, as shown in Fig \ref{relation_ratio}. There are significantly fewer nonedge sites in the configuration of quantum rings than in quantum dots, and there is an inner edge in quantum rings. The edge proportion of the TGQD is the highest of the three common quantum dots. Similarly, TGQR was studied because of its high proportion of edge regions compared to the total area.
We shall investigate only TGQRs with a hexagonal lattice of ring width, that is, the difference between the outer diameter and inner diameter of TGQRs in units of the graphene hexagonal lattice width. The potential differences in edge magnetism between quantum rings and quantum dots were also examined. As the ring width narrows, the ratio of edge atoms to the overall number of atoms increases.

\begin{figure}[tbp]
\includegraphics[scale=0.38]{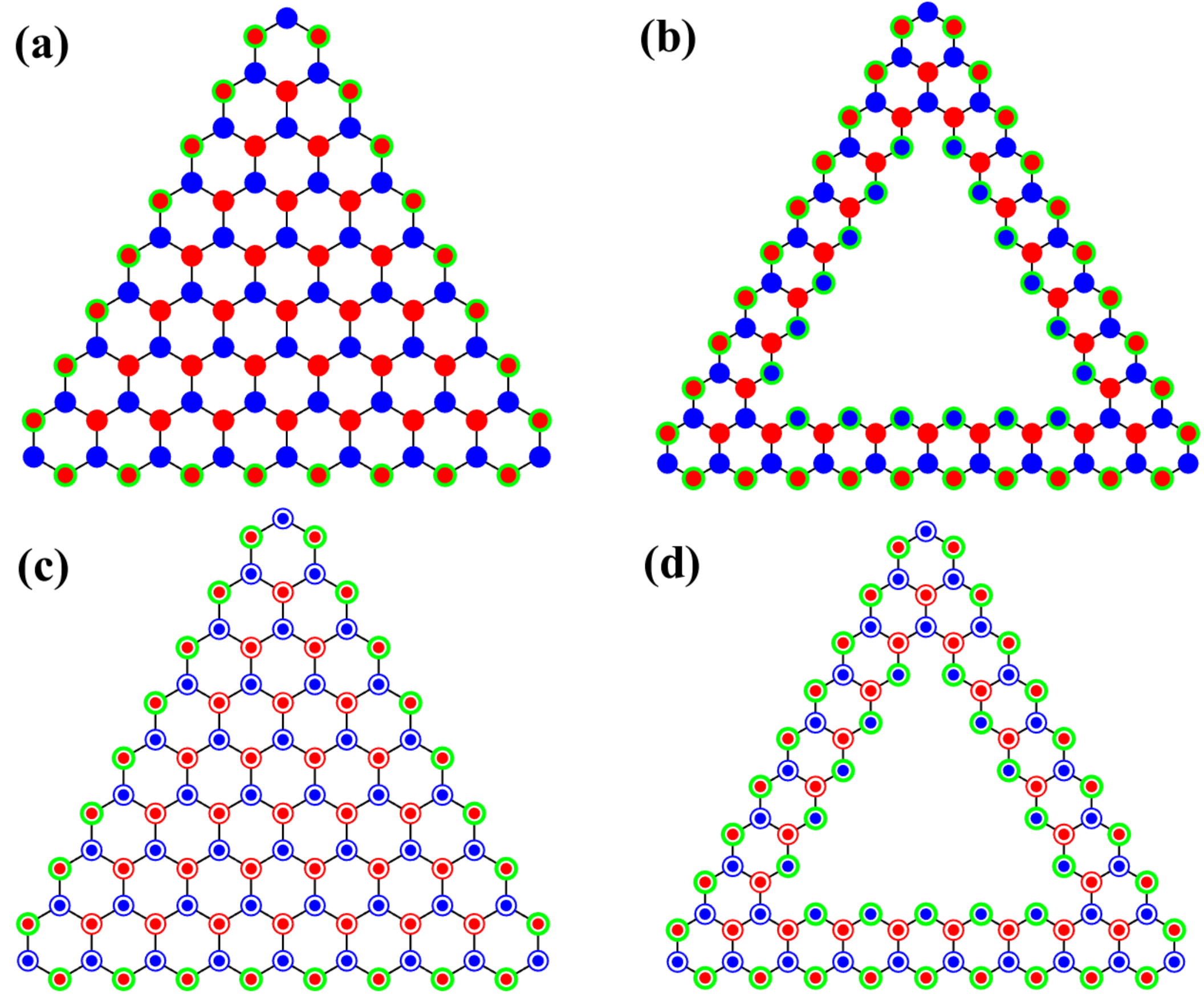}
\caption{ (Color online) Sketches for (a) a TGQD with 97 sites, (b) a TGQR with 105 sites, (c) a bilayer TGQD with 92 sites, and (d) a bilayer TGQR with 114 sites. In the subgraph (a) and (b), blue and red solid circles indicate A and B sublattices, respectively. In the subgraph (c) and (d), red and blue solid circles indicate A and B sublattices of the bottom layer, while red and blue hollow circles indicate sublattices of the top layer, respectively. The sites at the zigzag edge are marked by green marker edge.}
 \label{sketches}
\end{figure}

In Fig. \ref{sketches}(a), the TGQD sketch is presented, while Figs. \ref{sketches}(b)--\ref{sketches}(d) depict the TGQR, bilayer TGQD, and bilayer TGQR sketches.
In particular, solid circles stand for the sites of the top layer in the bilayer TGQD and the bilayer TGQR, and the hollow circles represent the bottom layer. A, B sublattices are distinguished by different colors, and the sites at the zigzag edge are highlighted by green marker edge.
The Hamiltonian of TGQD-like system is expressed as follows:
\begin{eqnarray}
H&=&H_k+H_k'+H_\mu+H_U.
\end{eqnarray}
Among them,
\begin{eqnarray}
H_k&=&-t\sum_{l	\left \langle i,j \right \rangle \sigma}(a^\dagger_{li\sigma}b_{lj\sigma}+\text{H.c.}),
\end{eqnarray}
\begin{eqnarray}
H_k'&=&-\sum_{i,j,l\neq l',\sigma}t_{ij}(a^\dagger_{li\sigma}a_{l'j\sigma}+a^\dagger_{li\sigma}b_{l'j\sigma}\nonumber\\
&+&b^\dagger_{li\sigma}a_{l'j\sigma}+b^\dagger_{li\sigma}b_{l'j\sigma}),
\end{eqnarray}
\begin{eqnarray}
H_{\mu}&=&\mu\sum_{i,l,\sigma}(n_{lai\sigma}+n_{lbi\sigma}),
\end{eqnarray}
\begin{eqnarray}
H_{U}&=&U\sum_{i,l}(n_{lai\uparrow}n_{lai\downarrow}+n_{lbi\uparrow}n_{lbi\downarrow}),
\end{eqnarray}
$H_k$ is the intralayer hopping term, and $H_k'$ is the interlayer hopping term, which is zero for monolayer graphene.
$H_\mu$ represents the chemical potential and $H_U$ represents the on-site Hubbard interaction.
$i$ and $j$ represent different lattice site index, and $\left \langle i,j \right \rangle$ represents a pair of nearest neighbors (NN). $a^\dagger_{li\sigma}(a_{li\sigma})$ creates (annihilates) electrons with spin $\sigma(\sigma=\uparrow,\downarrow)$ at the $i$ lattice point on sublattice A of the $l$ layer, as well as $b^\dagger_{li\sigma}(b_{li\sigma})$ acting on electrons of sublattice B. $n_{lai\sigma}=a^\dagger_{li\sigma}a_{li\sigma}$ and $n_{lbi\sigma}=b^\dagger_{li\sigma}b_{li\sigma}$.
$t$= 2.7 eV is the NN hopping integral,
which is a typical value that best reproduces the slopes of the valence and conduction bands at the $K$ point from DFT
calculations and is consistent with the experimental parameters \cite{RevModPhys.81.109,PhysRevB.66.035412,Andrei2020}. $t_{ij}$ is the hopping integral from lattice site $\boldsymbol{R_{1i}}$ of one layer to lattice site $\boldsymbol{R_{2j}}$ of the other layer, satisfying
\begin{equation}
t_{i j}=t_{\mathrm{c}} e^{-\left(\left|\boldsymbol{R}_{1 i}^d-\boldsymbol{R}_{2 j}^{d^{\prime}}\right|-d_0\right) / \xi}
\label{e6}.
\end{equation}
For $t_c=0.17t$, the vertical distance is $d_0=0.335$ nm, and $\xi=0.0453$ nm \cite{li2017splitting}.
It indicates the interlayer hopping from $\boldsymbol{R_{1i}}$ of the first layer to $\boldsymbol{R_{2j}}$ of the second layer,
which is related to the relative position of the two lattices $|\boldsymbol{R_{1i}}^d-\boldsymbol{R_{2j}}^{d'}|$.
$t_{ij}$ decreases exponentially with interlayer distance and becomes negligible beyond 3.0$a$.
A two-center Slater-Koster type model can be used to describe the $p_z$ orbitals on carbon atoms. The hopping terms are shown below \cite{PMID:20121163,moon2013optical}:
	\begin{equation}
	t(\boldsymbol{r})=V_{pp\pi}(\boldsymbol{r})+V_{pp\sigma}(\boldsymbol{r}),
		\label{equ7}
    \end{equation}
	where
	\begin{equation}
    V_{pp\pi}(\boldsymbol{r})=t_{pp\pi}e^{-\left(\left|\boldsymbol{R}_{1 i}^d-\boldsymbol{R}_{2 j}^{d^{\prime}}\right|-d_0\right) / \xi}
		\label{equ8}
	\end{equation}
	 and
	 \begin{equation}
	V_{pp\sigma}(\boldsymbol{r})=t_{pp\sigma}e^{-\left(\left|\boldsymbol{R}_{1 i}^d-\boldsymbol{R}_{2 j}^{d^{\prime}}\right|-d_0\right) / \xi}\left(\frac{\boldsymbol{r}\cdot\boldsymbol{e}_z}{|\boldsymbol{r}|}\right)^2. \label{equ9}
	 \end{equation}
	For Eq. (\ref{e6}), the parameter $t$ corresponds to the $pp\sigma$ term of the Slater-Koster hopping parameters, whereas $t_\mathrm{c}$ is associated with the $pp\pi$ term \cite{doi:10.1073/pnas.1108174108}.

We probe the edge magnetic properties of three different types of quantum dots (rings) at finite temperature using DQMC simulations.
In this method, the action $e^{-\beta H}$ is split into $M$ slices by Trotter decomposition, namely $e^{-\beta H}=\prod_{M}e^{-\Delta \tau H}$.
Then, the interaction term is decoupled by using Hubbard-Stratonovich transformation \cite{PhysRevB.40.506,PhysRevD.24.2278}.
These observations can be reproduced in calculation using a particular auxiliary field configuration because the action will be bilinear after transformation.
In practice, the target observations are obtained by sampling in the configuration space. The simulation provides 8000 warm-up sweeps to equilibrate the system, and 30000 sweeps were subsequently conducted for the measurements. The number of measurements was split into ten bins, which provide the basis for coarse-grain averages and errors estimated based on standard deviations from the averages so that the simulation can be performed at low enough temperatures to converge to the ground state.
At half filling, the particle-hole symmetry frees our system from the sign problem.

The uncertainty of the Coulomb interaction parameter $U$ in graphene and its derivatives is noteworthy. This value can be inferred from estimations made for polyacetylene \cite{herbutInteractionsPhaseTransitions2006,PhysRevLett.56.1509,RevModPhys.81.109}, where $U$ ranges between 6.0 eV and 17.0 eV , encompassing a broad range of values. To examine the influence of interactions on magnetic properties in such systems, our simulations consider $U$ values from 1.0$\vert t \vert$ to 4.0$\vert t \vert$.
In the remainder of this paper, we set $t$ as the unit.

\section{Results}
We introduce $\chi_b$ and $\chi_z$ to describe the magnetic susceptibility of the bulk quantum dot (ring) and the zigzag edge, respectively, in order to characterize the ferromagnetic behavior of the system \cite{PhysRevB.91.075410},
\begin{eqnarray}
\chi=\int_0^\beta d\tau \sum_{d,d'=a,b}\sum_{i,j}\langle m_{i_d}(\tau)\cdot m_{j_{d'}}(0)\rangle,
\end{eqnarray}
where $m_{i_a}(\tau)=e^{H\tau}m_{i_a}(0)e^{-H\tau}$, $m_{i_a}=a^\dagger_{i\uparrow}a_{i\uparrow}-a^\dagger_{i\downarrow}a_{i\downarrow}$, and sublattices A and B are equivalent. The bulk magnetic susceptibility $\chi_b$ is the average of the $zz$ spin correlation of all sublattices, and $\chi_z$ is the average of the sublattice of the zigzag edge of the quantum dot (ring), which is shown in Fig. \ref{sketches} with special marking edges.

\begin{figure}[tbp]
\includegraphics[scale=0.36]{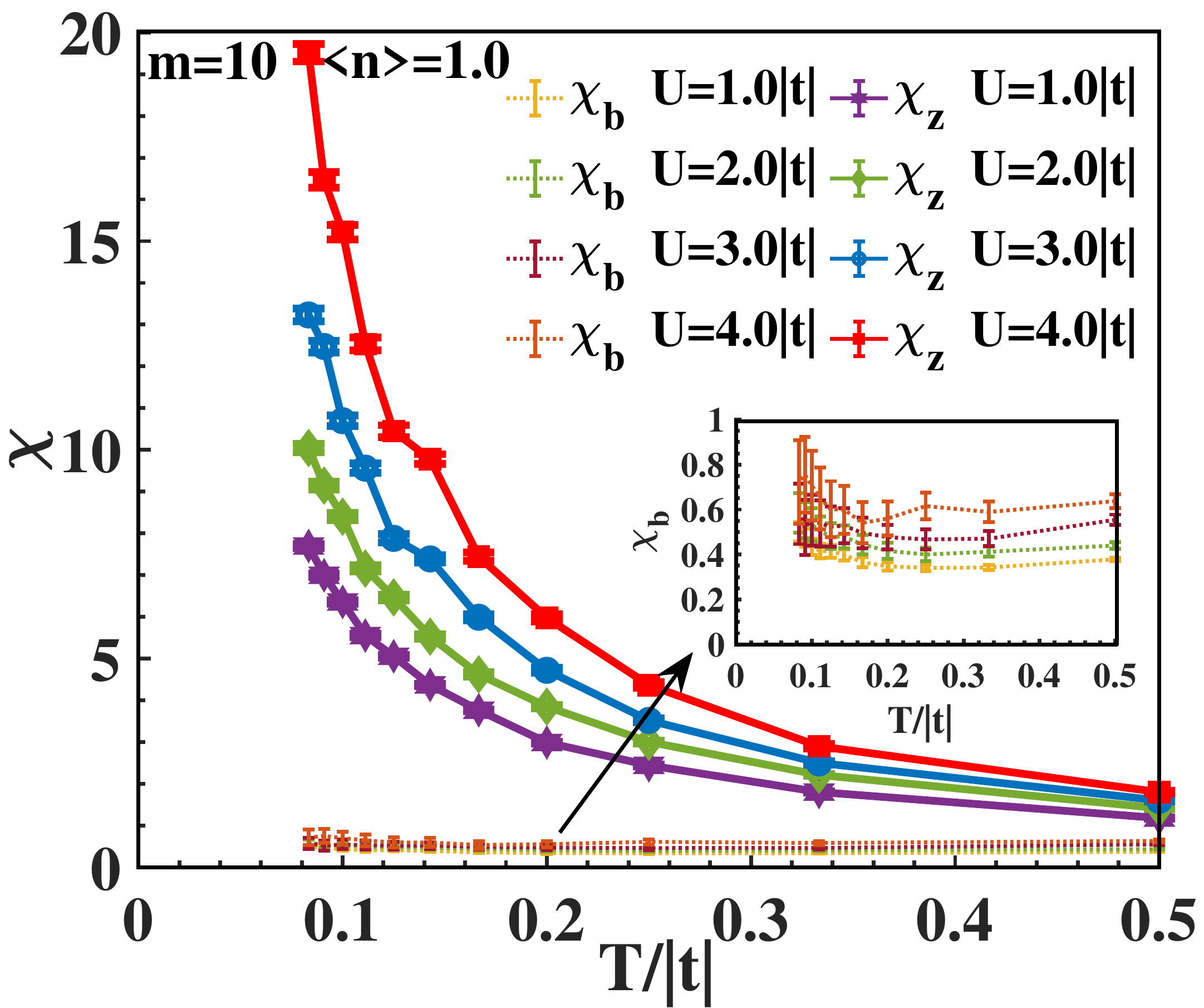}
\caption{ (Color online) The $\chi_z$ (solid line) and $\chi_b$ (dotted line) of a TGQR at $\langle n\rangle=1.0$ with different $U$, and $m$ in the tag represents the number of hexagons contained by the edge of the triangle ring.
} \label{chi_U}
\end{figure}

To determine how the edge magnetic susceptibility of TGQRs changes with temperature at various on-site interactions $U=1.0 \vert t \vert \sim4.0 \vert t \vert$,
we first plot Fig. \ref{chi_U} and that $\chi_z$ decreases with temperature in an inversely proportional way. We assume that the edge of the TGQR is ferromagnetic according to the Curie-Weiss law $\chi=C/(T-T_C)$.
The law shows the connection between magnetic susceptibility and temperature in ferromagnetic materials below the Curie temperature $T_C$.

\begin{figure}[tbp]
\includegraphics[scale=0.385]{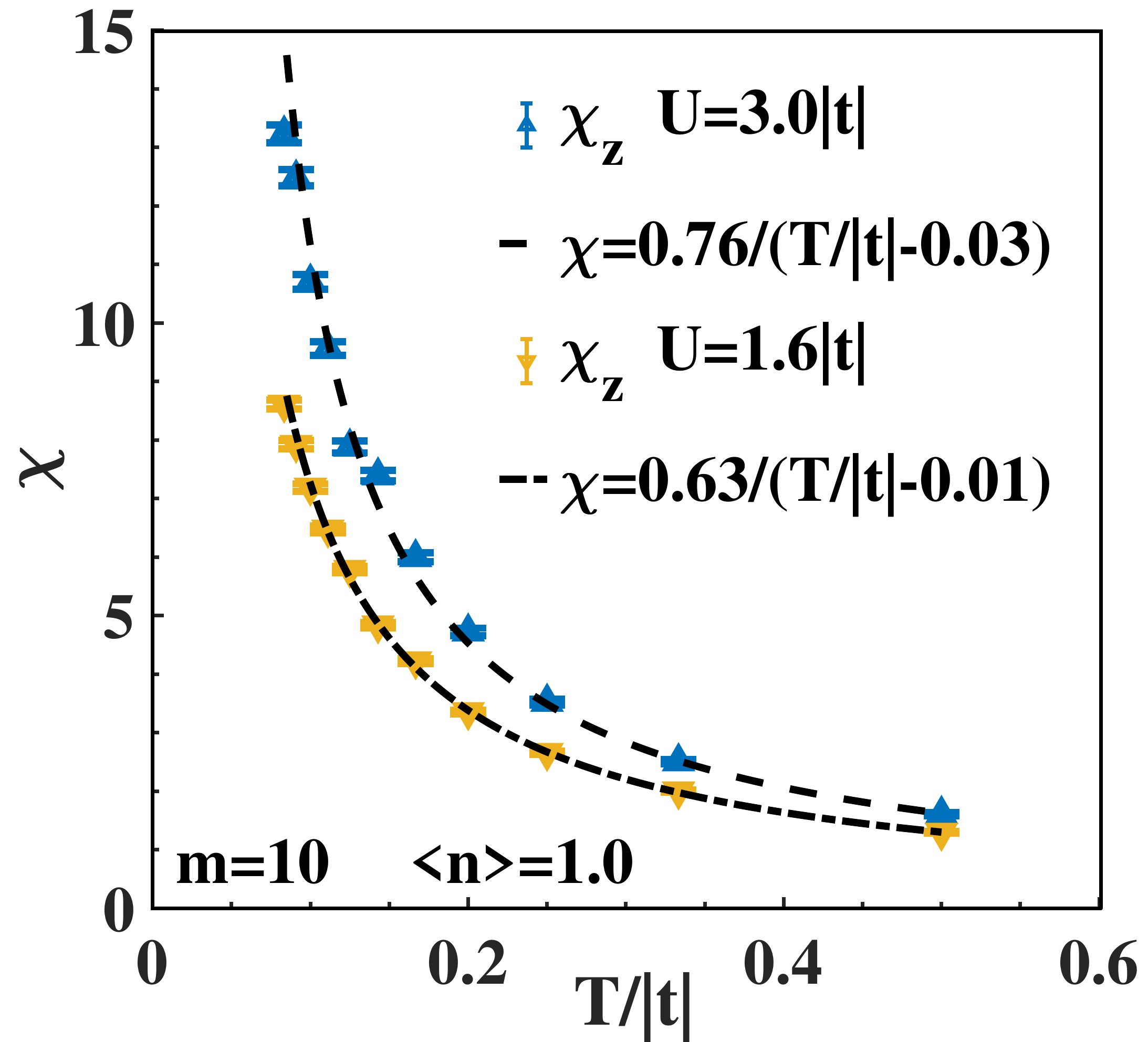}
\caption{ (Color online) The $\chi_z$ of a TGQR of $m=10$ at $\langle n\rangle=1.0$ with $U=1.6 \vert t \vert$ and $3.0 \vert t \vert$ and the fitting of Curie's law, respectively.
} \label{chi_fitting}
\end{figure}

We choose two values of $U$ for fitting.
The first typical $U$ is derived from the Peierls-Feynman-Bogoliubov variational principle,
which maps a generalized Hubbard model with nonlocal Coulomb interactions onto an effective Hubbard model with only on-site effective interactions $U$, demonstrating that $U$ is approximately $1.6 \vert t \vert$ \cite{PhysRevLett.111.036601}.
The second one, $U=3.0 \vert t \vert$ \cite{PhysRevB.91.075410}, is a typical value for examining how interactions affect the magnetic characteristics of quantum dots.
For these calculations, we use the following formula:
\begin{eqnarray}
\chi=\frac{C}{T-T_C}.
\end{eqnarray}
According to Fig. \ref{chi_fitting}, TGQR exhibits conventional ferromagnetic behavior, and the relationship between $\chi_z$ and temperature satisfies the Curie-Weiss law.
The $T_C$ is approximately $0.033\vert t \vert$ at $U=3.0 \vert t \vert$ and approximately $0.013\vert t \vert$ at $U=1.6 \vert t \vert$.

The Monte Carlo approach results in larger intrinsic variances due to the process of sampling at lower temperatures, which produces a slightly incorrect result. The magnetic susceptibility at the lowest temperature in the calculation result is used to estimate the maximum inaccuracy of $T_C$,
\begin{eqnarray}
\delta T_C=\frac{C}{\chi^2}\delta\chi.
\end{eqnarray}
For $U=3.0 \vert t \vert$, $\delta T_C$ is approximately $0.008\vert t \vert$. For $U=1.6 \vert t \vert$, $\delta T_C$ is approximately $0.005\vert t \vert$.
Our results demonstrate the edge ferromagnetic feature even when the interaction $U$ is small.

The bulk magnetic susceptibility $\chi_b$ is flat and can be deduced from the half-filled Hubbard model on the ideal honeycomb lattice due to its antiferromagnetism in the ground state.

For any central atom, the NN atom has a negative spin correlation factor because the hexagonal honeycomb lattice is antiferromagnetic. This spin correlation between the next-nearest neighbor (NNN) atom and the central atom is positive, that is, ferromagnetic correlation, because the NNN atom and NN atom have a NN relationship. The center atom and the NNN atom are connected by a NN relationship to the same sublattice, according to the graphene structure. Therein, the high dependence of $\chi_z$ on temperature may be caused by the fact that the atoms at the zigzag edges belong to the same sublattice.\par

\begin{figure}[tbp]
\includegraphics[scale=0.38]{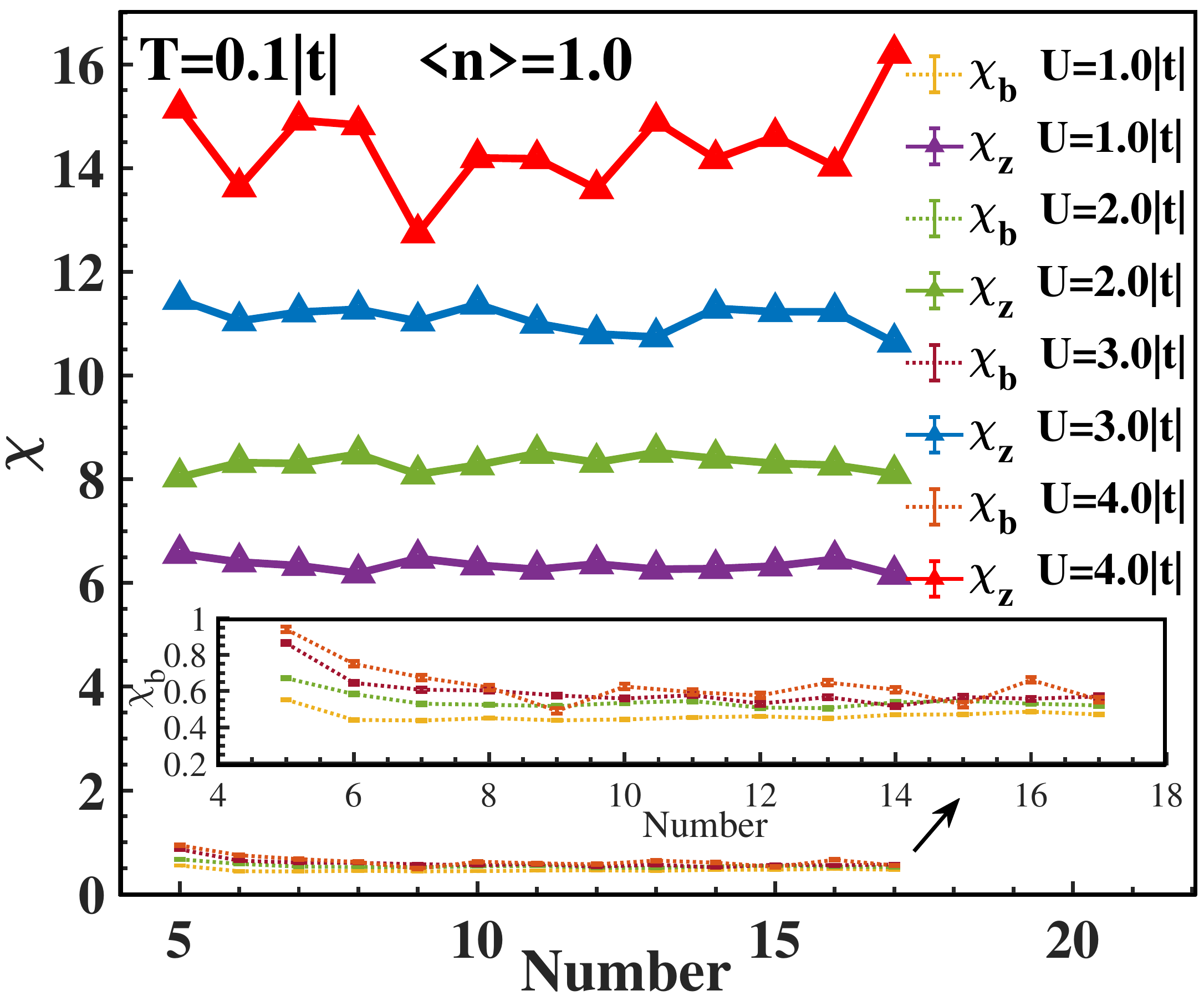}
\caption{ (Color online) The $\chi_z$ (solid line) and $\chi_b$ (dotted line) of TGQR of different size at $\langle n\rangle=1.0$ with different $U$. ``Number" represents for the number of hexagons contained by the edge of the triangle ring.
}\label{chi_size}
\end{figure}

We show $\chi_b$ and $\chi_z$ as the number of hexagonal lattices at the TGQR edge shifts from 5 to 17 under the conditions of $T=0.1\vert t \vert$ and $\langle n\rangle=1.0$ to investigate the impact of the size effect on TGQR edge ferromagnetism, as shown in Fig. \ref{chi_size}. With an increase in the quantum ring size, $\chi_b$ and $\chi_z$ gradually change. The magnetic susceptibility scarcely changes with structure size, especially at low $U$ ($U=3.0 \vert t \vert$ and below), suggesting that the robustness of TGQR with zigzag edges is unaffected by size. This phenomenon occurs because the TGQR edge structure is a typical zigzag structure and has a significant ferromagnetic correlation. In addition, we deduce that a larger on-site interaction $U$ can greatly increase the Curie temperature of the system by enhancing the magnetic susceptibility of the edge.\par

Starting with the smallest TGQR, each side of the triangle contains five honeycomb lattices, that is, five zigzag edge atoms. In this case, the ferromagnetic correlation of its edge has been saturated, signifying that the spins of the electrons on the edge are practically in the same direction. As a result, adding more atoms with zigzag edges will not strengthen the ferromagnetic correlation. A change in system size may result in an erratic oscillation of the edge magnetic susceptibility at higher $U$. At the same temperature, we also observe from Figs. \ref{chi_U} and \ref{chi_size} that $\chi_b$ and $\chi_z$ increase as $U$ increases.
Figure \ref{chi_fitting} indicates that as $U$ increases, the Curie temperature increases.

\begin{figure}[tbp]
\includegraphics[scale=0.42]{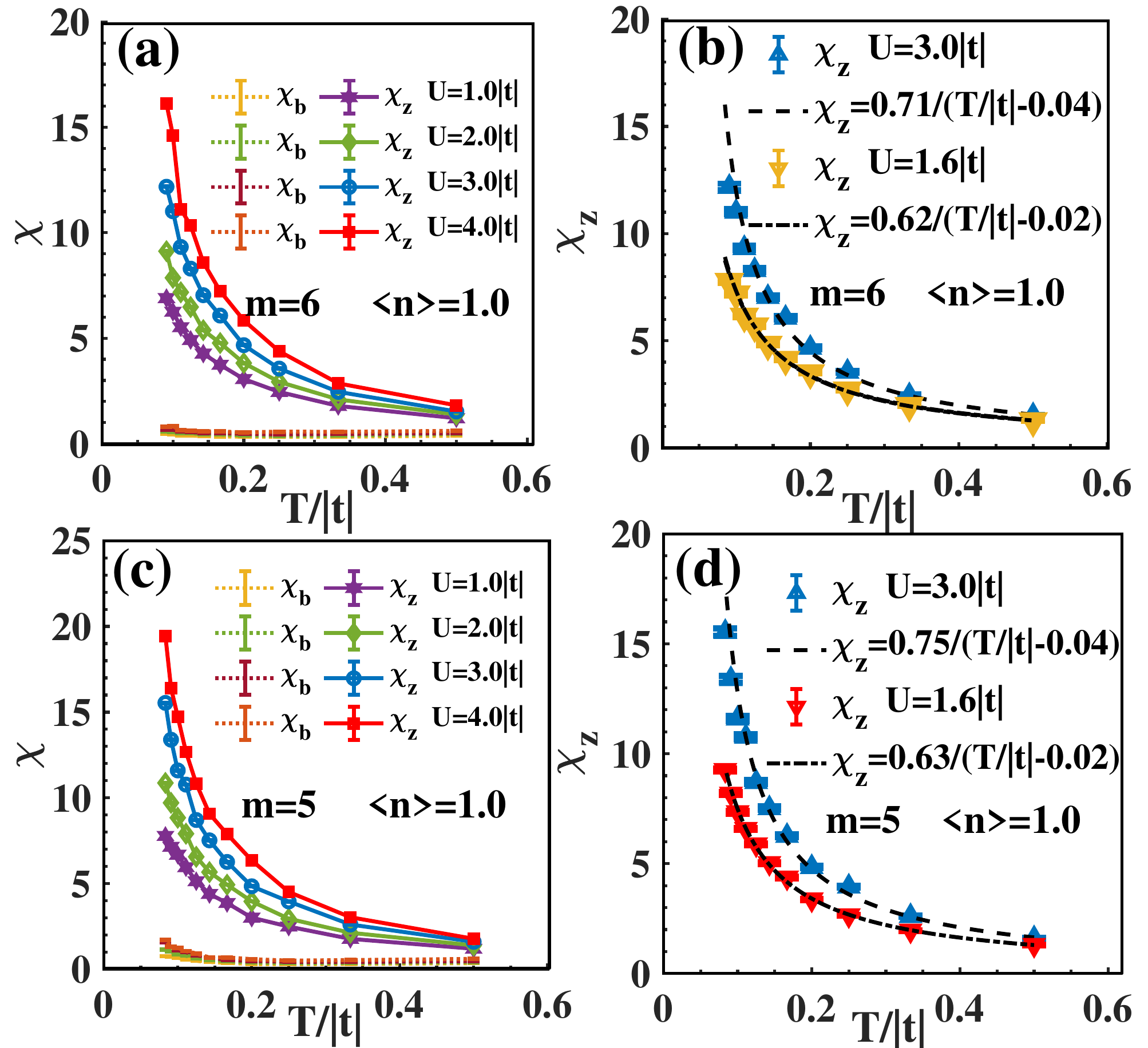}
\caption{(Color online) $m$ in the tag represents the number of hexagons contained by the edge of (a) bilayer TGQR/ (c) bilayer TGQD. The $\chi_z$ (solid line) and $\chi_b$ (dotted line) of (a) a bilayer TGQR with 114 sites and (c) a bilayer TGQD with 92 sites at $\langle n\rangle=1.0$ with $U=1.0\sim4.0 \vert t \vert$. The $\chi_z$ of (b) a bilayer TGQR and (d) a bilayer TGQD with $U=1.6 \vert t \vert$ (positive triangle marker) and $3.0 \vert t \vert$ (inverted triangle marker), the dashed line represents the fitting of Curie's law when $U=3.0 \vert t \vert$, while the dotted line represents the fitting when $U=1.6 \vert t \vert$.}
\label{six}
\end{figure}

\begin{figure}[tbp]
\includegraphics[scale=0.22]{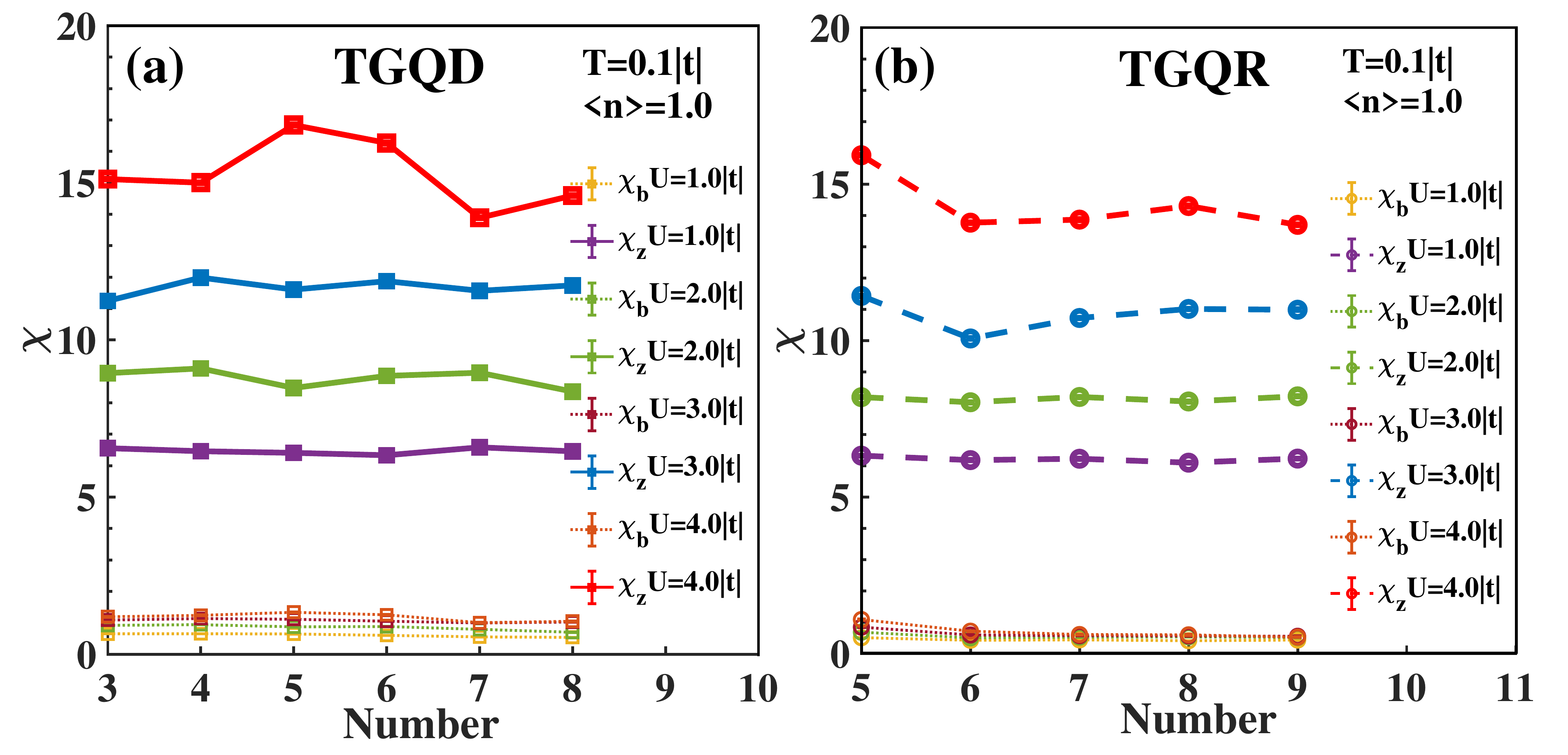}
\caption{(Color online) The $\chi_z$ and $\chi_b$ of bilayer TGQD (square marker) and bilayer TGQR (circle marker) of different size as shown in the subgraphs with different $U$. ``Number" represents the number of hexagons contained by the edge of bilayer TGQD or bilayer TGQR.}
\label{seven}
\end{figure}

We estimated the changes in magnetic susceptibility of TGQD and TGQR bilayers with temperature and size to investigate the impact of interlayer coupling on the edge magnetism.
The DQMC can be applied to approximately 100 quantum dot (ring) lattice sites with precise sampling and calculation of physical quantities, and the result is shown in Fig. \ref{six}. In the half-filled system, the $\chi_z$ and $\chi_b$ of the bilayer TGQD and the bilayer TGQR increase with increasing $U$, and $\chi_z$ dramatically decreases with increasing $T$; however, the change in $\chi_b$ with $T$ is comparatively flat.

In the quantum ring, the inner edge atoms and the outer edge atoms do not belong to the same sublattice, and the value of their spin correlation is likely to be negative.
When we calculate the edge magnetic susceptibility of the quantum ring, the spin correlations of the inner and outer boundaries are separately calculated and then these values are subsequently added.
Thus, the edge magnetism is not directly impacted by the spin correlation between the inner and outer edges.

As illustrated in Figs. \ref{six}(b) and (d), $\chi_z$ with respect to temperature is fitted under the on-site interactions $U=3.0\vert t \vert$ and $U=1.6\vert t \vert$.
The Curie-Weiss law is satisfied by the edge magnetic susceptibilities of bilayer
TGQDs and bilayer TGQRs,
and there is a strong edge ferromagnetic correlation.
This reveals that the zigzag edge of the bilayer TGQD
has a slightly greater Curie constant $(C)$ than the zigzag edge of the bilayer TGQR.
The former is $U=3.0\vert t \vert$, $C=0.75$ and $U=1.6\vert t \vert$, $C=0.63$.
The latter is $U=3.0\vert t \vert$, $C=0.71$ and $U=1.6\vert t \vert$, $C=0.62$, and their critical temperatures for the ferromagnetic-paramagnetic phase transition are similar.
The reason might be that the bilayer TGQD already has significant edge ferromagnetism,
and the ring structure adds more ferromagnetic boundaries while the original edge ferromagnetism remains the same.
Because relatively few zigzag edge atoms have already demonstrated
a very strong ferromagnetic correlation,
as shown in Fig. \ref{seven}, $\chi_z$ and $\chi_b$ of the bilayer
TGQD and bilayer TGQR is nearly unchanged with the increase of structure size.

\section{Summary of results}
In summary, based on the Hubbard model, we computed the robust edge ferromagnetism of the TGQD-like system using DQMC.
The TGQD-like structure, which includes TGQRs, bilayer TGQDs, and bilayer TGQRs, was chosen because it reflects the zigzag border condition among the various quantum dot (ring) shapes. Our results show that temperature has a significant impact on the edge magnetic susceptibility of such systems. We use the Curie-Weiss law to fit the TGQD-like system, which shows that it exhibits robust edge ferromagnetic behavior. One way to improve the edge magnetic susceptibility and increase the Curie temperature is to increase on-site Hubbard interactions $U$. The application of TGQD-like systems in spintronics may benefit from robust edge ferromagnetic behavior.\par

\noindent
\underline{\it Acknowledgments}
This work was supported by Beijing Natural Science
Foundation (No. 1242022), NSFC (No. 11974049), and
Guangxi Key Laboratory of Precision Navigation Technology and Application,
Guilin University of Electronic Technology (No. DH202322).
The numerical simulations in this work were performed at the HSCC of Beijing Normal University.

\section*{appendix}
\begin{figure}[tbp]
\includegraphics[scale=0.32]{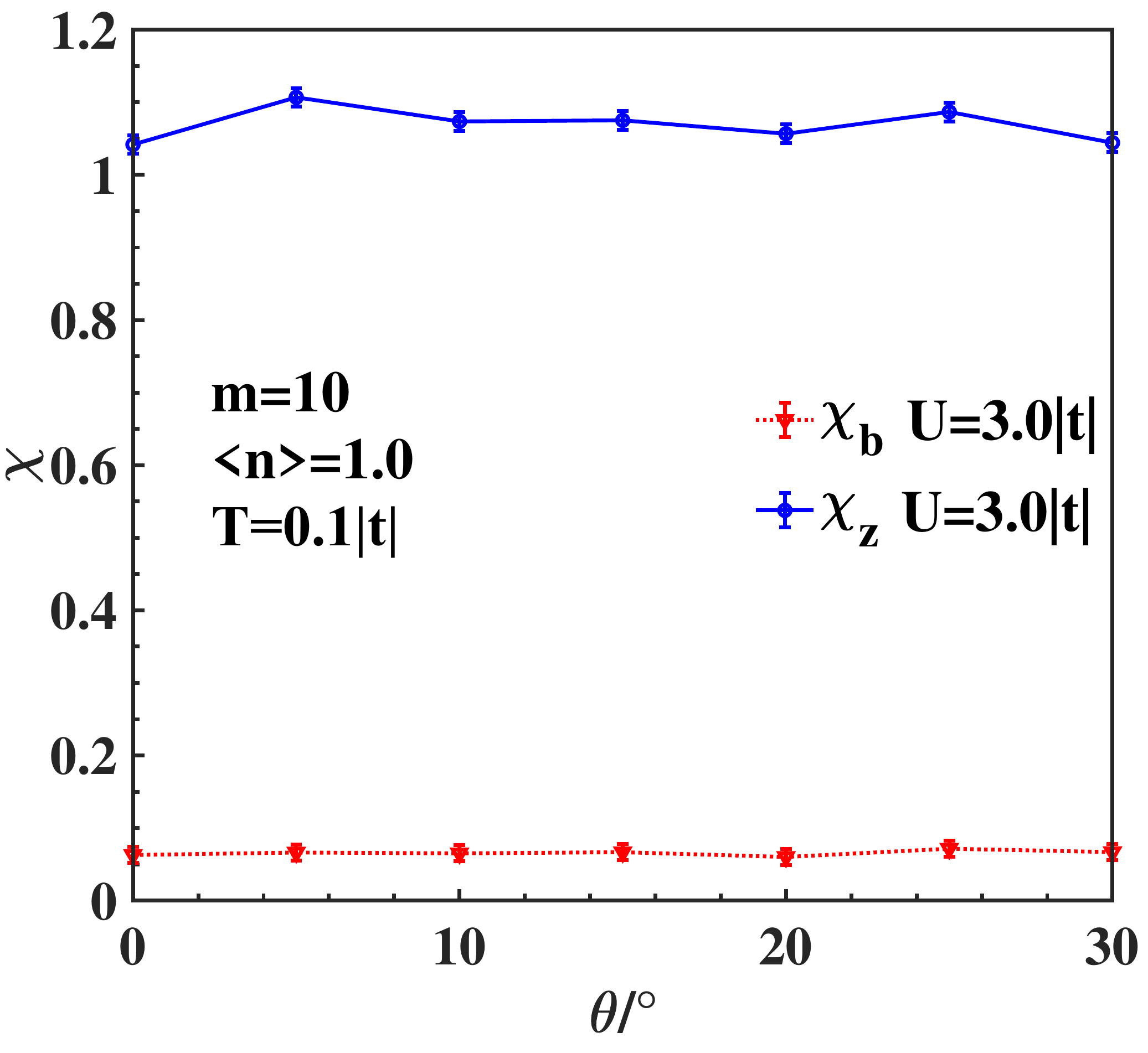}
\caption{ (Color online) The ferromagnetic susceptibility of a bilayer TGQR in various twist angles.}
\label{Fig8}
\end{figure}

\begin{figure}[tbp]
\includegraphics[scale=0.32]{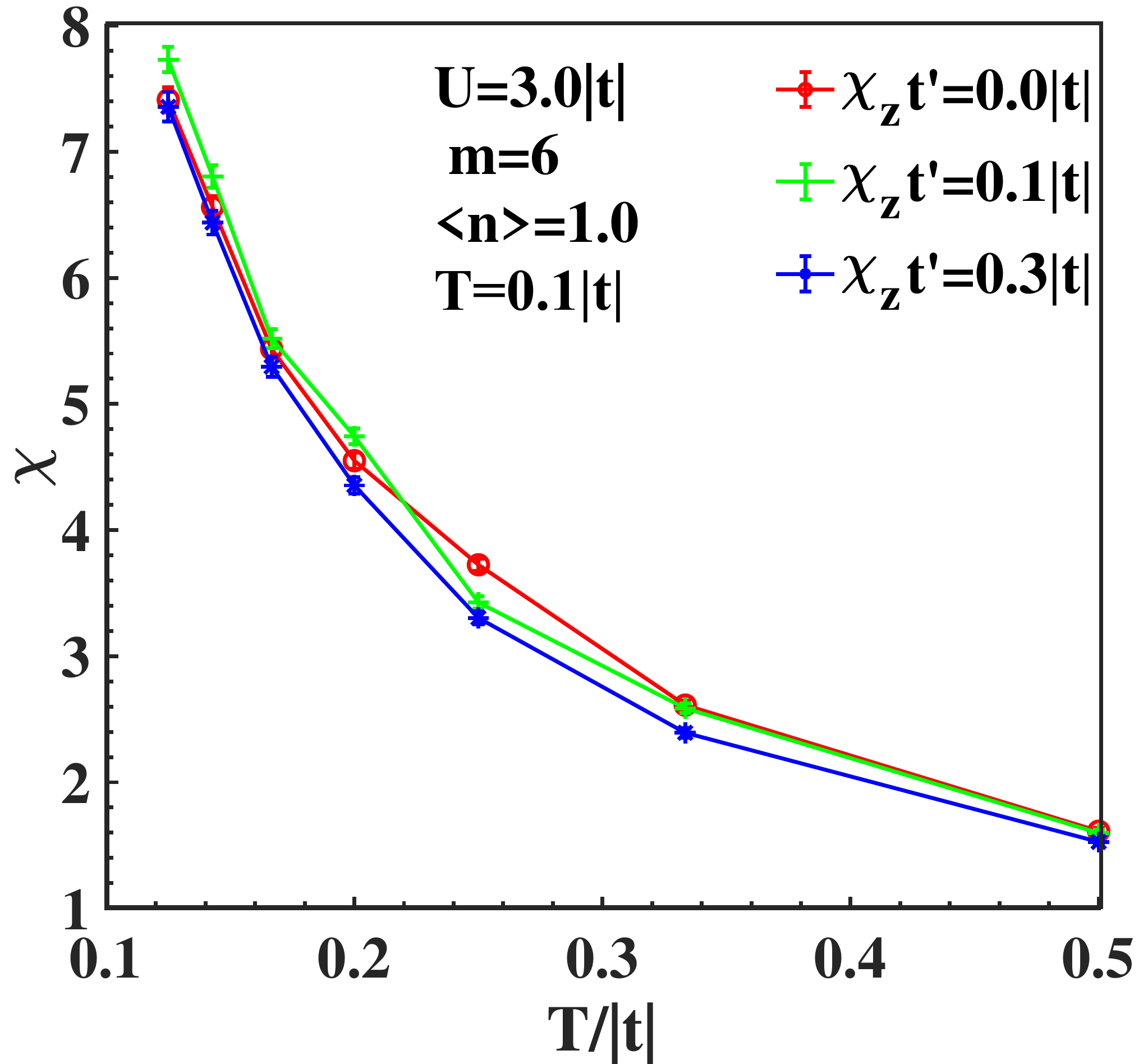}
\caption{ (Color online)
The ferromagnetic susceptibility as a function of temperature at the zigzag edge when NNN hopping is introduced.
}\label{Fig9}
\end{figure}

When simulating the physics of many-body interaction systems, the computational cost grows exponentially with the size of the lattice. Therefore, our simulations are confined to a relatively small system and may provide some enlightening results for describing truncated small lattices.

We provide the results for different angles below to validate our study for the lattice size we simulated.
We can see that the magnetic
susceptibility hardly changes with the twist angle in Fig.~\ref{Fig8}.

\begin{figure}[tbp]
\includegraphics[scale=0.23]{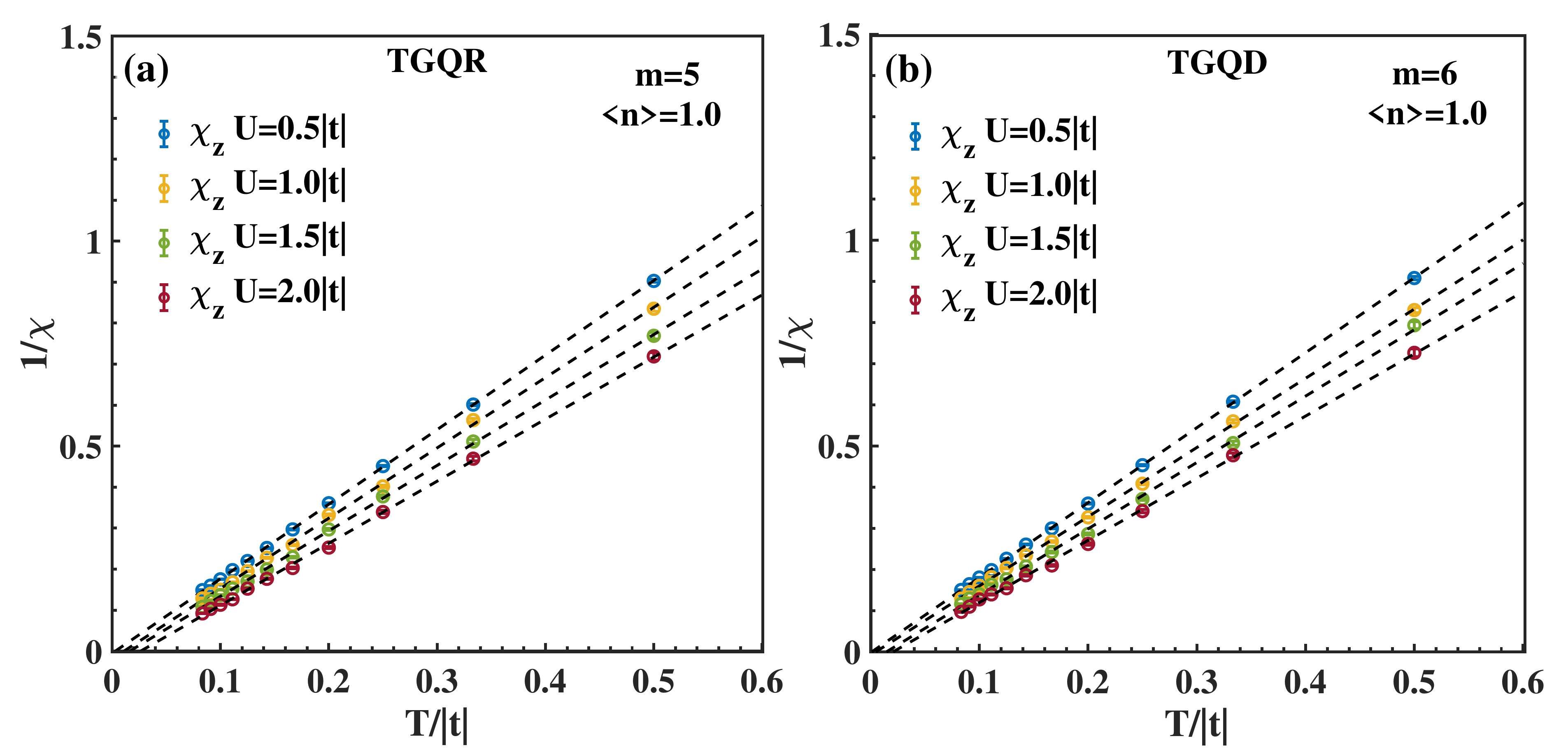}
\caption{(Color online) Linear fitting of the inverse of $\chi_z$ of TGQR and TGQD versus temperature at half-filling state.}
\label{Fig10}
\end{figure}

We next discuss different NNN hopping parameters in the TGQRs in Fig.~\ref{Fig9}.
Our results show that ferromagnetism is not affected by
NNN hopping at $t^{\prime}=0.0 \vert t \vert, 0.1 \vert t \vert, 0.3 \vert t \vert$,
which is consistent with the analytic expression for tight-binding dispersion \cite{PhysRevB.66.035412}.
Therefore,
our simulations show that this system has robust edge ferromagnetism even when considering
intralayer NNN hopping.

In order to further investigate the effect of the on-site interaction $U$ on the Curie temperature,
we plot the reciprocal of $\chi$ as depicted below in Fig.~\ref{Fig10}.
The figures exhibit a linear correlation between $1/\chi$ and temperature $T$,
which corresponds to the Curie-Weiss behavior $1/\chi = (T-T_c)/A$.
Therefore, we extrapolate $1/\chi$ to zero temperature by using linear fitting.
If the system possesses a finite $T_c$, its intercept should be negative. Figure~\ref{Fig10}
shows that the on-site interaction $U$ enhances the ferromagnetism and
the absolute value of the intercept becomes larger with increasing $U$ both in TGQR and TGQD.\par

\begin{figure}[tbp]
\includegraphics[scale=0.23]{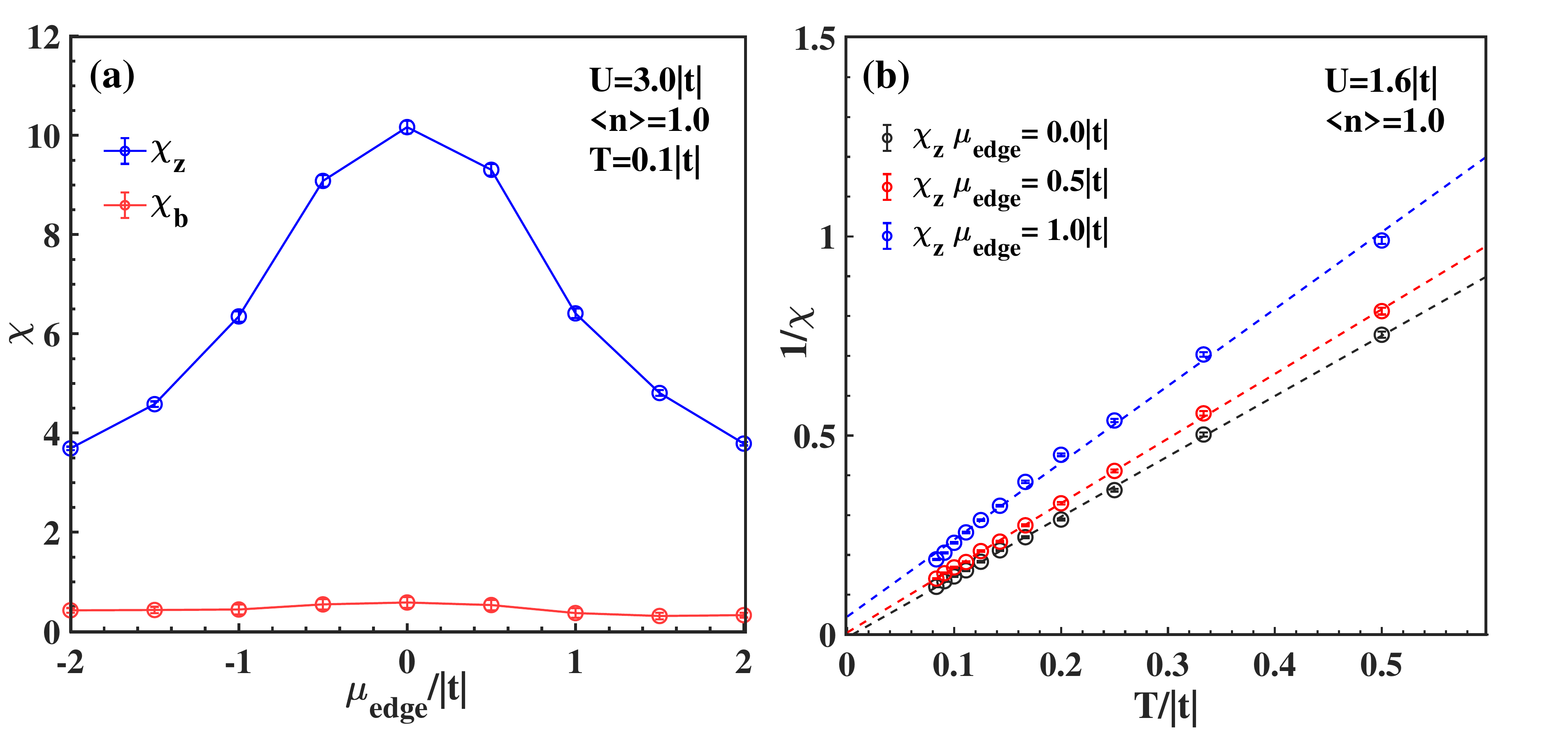}
\caption{(Color online) (a) Linear fitting of the inverse of $\chi_z$ of TGQR versus temperature at different edge chemical potentials and the whole system maintains half-filling state. (b) Ferromagnetic susceptibility $\chi_z$ and $\chi_b$ of TGQR at 114 sites for different boundary chemical potentials at $\langle n \rangle=1.0$, and $\beta=1/T$ represents the reciprocal of the temperature.}
\label{Fig11}
\end{figure}

We further add an additional potential at the zigzag edges of the system,
to investigate a realistic system in which zigzag edges are decorated by adding atoms. To accomplish this, we change the chemical
potential of the edge atoms from $-2 \vert t \vert$ to $2 \vert t \vert$ and adjust the bulk chemical potential to maintain half-filling, as shown in Fig.~\ref{Fig11}(a). Our simulation results show that the magnetic susceptibility $\chi_z$ has a maximum value when the chemical potential of the edge atoms is 0, and $\chi_b$ remains almost stable in this range. Additionally, Fig.~\ref{Fig11}(b) indicates that as the absolute value of the edge potential changes from 0.0 to $0.5|t|$ at half-filling state, the edge ferromagnetism of the system gradually disappears.
It is interesting that even a tiny $U$ is enough to make the edge
ferromagnetic at half filling.  If including a nonzero edge potential, this phenomenon disappears and one needs finite $U$ for magnetism to appear where edge states getting nonflat dispersion. This behavior could be expected from Stoner criterion. Thus, our results provide important guidance for understanding the physics of TGQD-like system edge doping experiments.

\begin{figure}[tbp]
\includegraphics[scale=0.36]{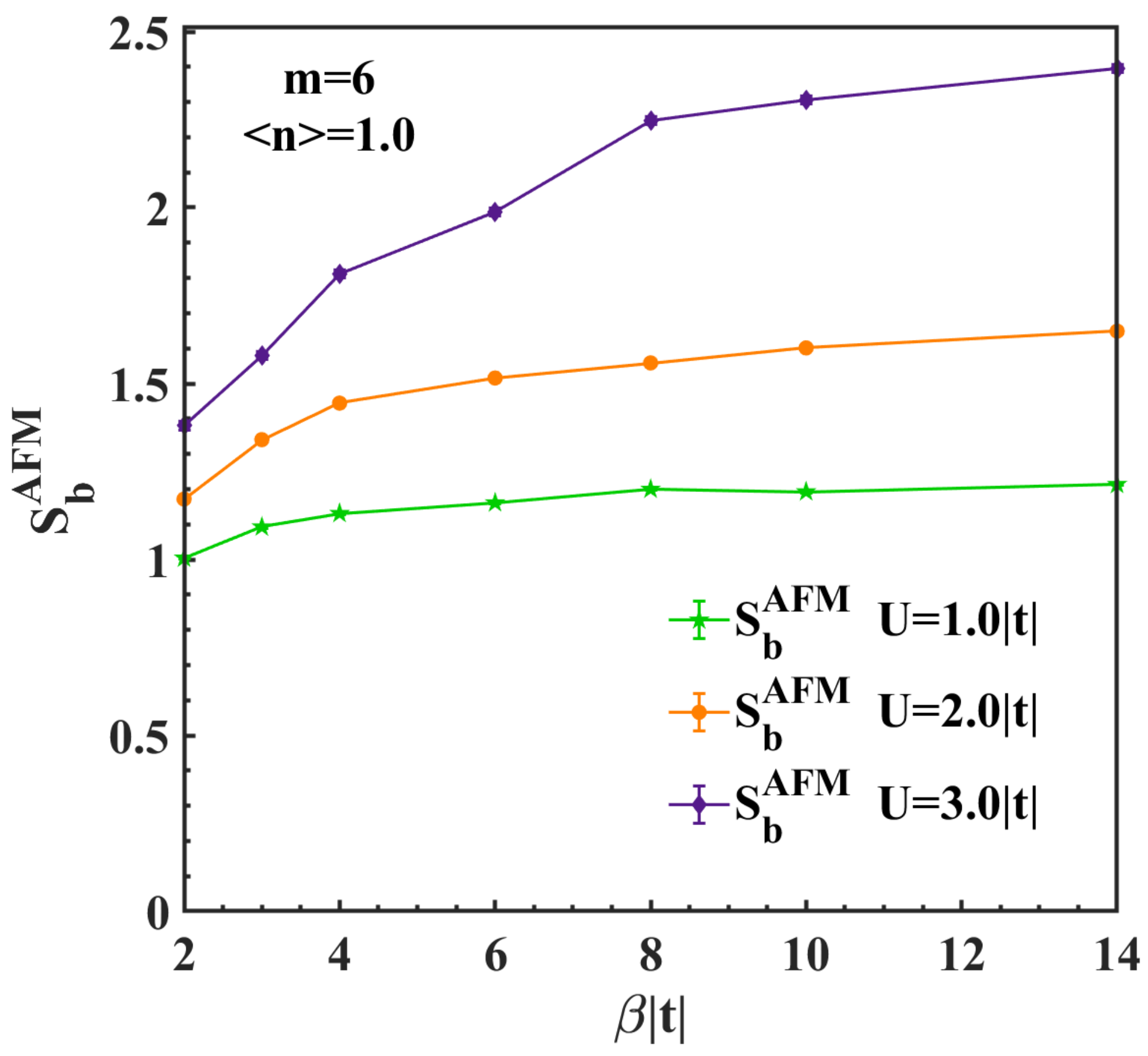}
\caption{(Color online) The antiferromagnetic correlation of bilayer TGQR as a function of $\beta=1/T$ for different $U = 1.0\vert t \vert \sim 3.0\vert t \vert $.}
\label{Fig12}
\end{figure}

To inspect the validity of inverse temperature $\beta$ chosen in the main text, we refer to Fig.~\ref{Fig12}. Because the bilayer TGQR is the most complex structure among those discussed in our work, we consider this material as an example worth further discussion. The values of the correlation functions $S_b^{AFM}$ tend to stabilize with the increase of $\beta$ at $U = 1.0 \vert t \vert \sim 3.0 \vert t \vert$, where $S_b^{\mathrm{AFM}}=\frac{1}{N_\mathrm{{t}}}\left\langle\left[\sum_{l i}\left(\widehat{S}_{lai}^z-\widehat{S}_{lbi}^z\right)\right]^2\right\rangle$ \cite{HUANG2019310}, and $\widehat{S}_{lai}^z$ ($\widehat{S}_{lbi}^z$) is the $z$ component spin operator on A (B) sublattice of layer $l$. So the value of temperature used in our paper is sufficiently large.

\bibliography{ref}

\end{document}